\documentclass[a4paper, preprint]{jpconf}
\usepackage{graphicx}
\usepackage{amssymb}
\usepackage{amsmath}
\usepackage{amsfonts}
\begin{document}
\title{Locking-time and Information Capacity in CML with Statistical Periodicity}
\author{R M Szmoski, R F Pereira, F A Ferrari, S E de S Pinto}

\address{Departamento de F\'isica, Universidade Estadual de Ponta Grossa, 84030-900, Ponta Grossa, PR, Brazil.}

\ead{desouzapinto@pq.cnpq.br}

\begin{abstract}
In this work we address the statistical periodicity phenomenon on a coupled map lattice. The study was done based on the asymptotic binary patterns. The  pattern multiplicity gives us the lattice information capacity, while the entropy rate allows us to calculate the locking-time. Our results suggest that the lattice has low locking-time and high capacity information when the coupling is weak. This is the condition for the system to reproduce  a kind of behavior observed in neural networks.
\end{abstract}

\section{Introduction}

There are low-dimensional dynamical systems whose density evolution does not converge to an invariant state. Instead, it becomes periodic in time with well defined oscillation period. Such behavior is called statistical periodicity \cite{B-Mackey}. Dynamical systems with statistical periodicity, although possess an invariant density, do not evolve to it \cite{PhysicaD53,JSP63}. Once in physical world the asymptotic equilibrium is rather an exception than the rule, systems with statistical periodicity are essentials for understanding nonequilibrium phenomena. 

One-dimensional maps, such as the logistic and tent ones, exhibit statistical periodicity for certain values of the control parameter. For tent map it was demonstrated \cite{PhysicaD53} there is statistical periodicity for $(\sqrt{2})^{\frac{1}{T}}<a\le (\sqrt{2})^{\frac{1}{T-1}}$, where  $T=1,2,\cdots$ is the oscillation period and $a$ the map control parameter. In general, both logistic map and tent one exhibit statistical periodicity just at the parameter values where they generate banded chaos or quasiperiodicity.

In this work, we study a coupled map lattice (CML) composed by $N$ maps with statistical periodicity. CMLs are suitable mathematical models for spatially extended systems and spatiotemporal chaos \cite{B-kaneko}. Furthermore, they are considered useful to investigate neural networks \cite{PRL92,JP94}, because they can reproduce some experimentally observed features such as: a rapid response to stimuli, an irregular or chaotic behavior of individual elements and a great variability of patterns.   

The dynamic of a CML is, in general, governed by two competing mechanisms: a local reaction, which is given by map, and interaction due to coupling. Here, using the local-type coupling we shall try to answer two questions: a) How does the coupling  influence the asymptotic behavior? b) Is there any value interval of the coupling parameter from which the statistical periodicity is no longer observed? 

CMLs with statistical periodicity were studied in \cite{JP94,PRE67} driven by experimentally observed features in neural network. Hauptmann {\it et al} \cite{PRE67} show that the CML can reproduce such features when the coupling is weak. However, they determined the time for the lattice to achieve the permanent periodic state or locking-time using a somewhat primitive method. Here, we propose an accurate method to determine such quantity, which is based on the system entropy rate. Conversely, we show that there exist a dependence between locking-time and accessibility of effective binary states. Also we calculate the conditional entropy as information capacity measure of the CML. This paper is organized as follows: section 2 describes the coupling-type and map used; section 3 presents the periodicity detection method proposed; section 4 the findings and discussions and, finally, the conclusions.

\section{The lattice}
 We consider that system state at each instant of time $n$ is completely defined by the vector ${\bf x}_n=(x^{(1)}_n,x_n^{(2)},\cdots,x_n^{(N)})^T$, in which the components evolve according to the equation  

\begin{eqnarray}
x_{n+1}^{(i)}=(1-\varepsilon)f(x_n^{(i)})+\frac{\varepsilon}{2}\left[f(x_n^{(i-1)})+f(x_n^{(i+1)})\right],
\label{rede}
\end{eqnarray}
where $\varepsilon\in[0,1]$  is the coupling strength between the sites of the lattice and $f(x)$ is local dynamics. 

\begin{figure}[h]
\begin{minipage}{16pc}
\centering
\includegraphics[width=15pc,angle=-90,clip]{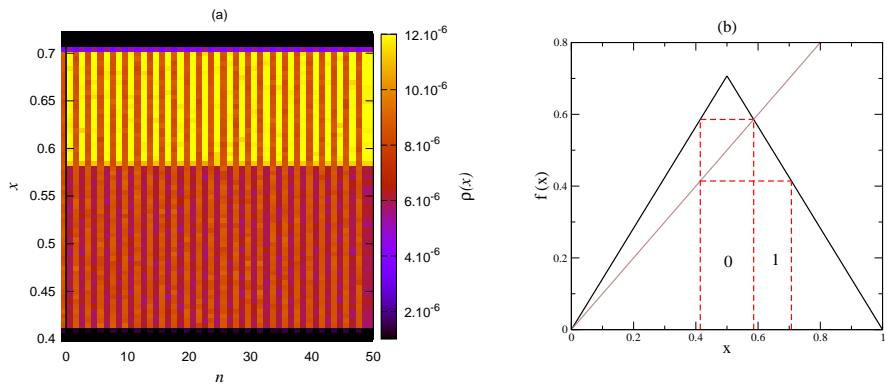}
\end{minipage}\hspace{3pc}%
\begin{minipage}{14pc}
\centering
\includegraphics[width=11pc,angle=0,clip]{fig2.eps}
\end{minipage} 
\caption{\label{Fig1} (a) Density evolution for an ensemble of $10^5$ uncoupled tend maps with $a=\sqrt{2}$. The states (time) correspond to vertical (horizontal) axis while the colors code the state density. Color intense (yellow online) corresponds to higher density, while dark color (red online) lower density. (b) $F_a(x)$ on intervals $[0,1]$ for $a=\sqrt{2}$ and partition generated by mapping the density boundaries.}
\end{figure}

Following \cite{PRE67} we consider
\begin{eqnarray}
f(x)=\left\{
\begin{array}{ccl}
ax &,& 0\le x < \frac{1}{2}\\
a(1-x)&,& \frac{1}{2}\le x\le 1\end{array}\right.,
\label{mapa}
\end{eqnarray}
with $a=\sqrt{2}$. For this parameter value the density flow exhibits a period-2 oscillation in time. Figure \ref{Fig1}(a) shows the evolution of a ensemble with  $100,000$ maps of the form   (\ref{mapa}), in which $a=\sqrt{2}$ and $T=2$. This periodicity can be viewed as the alternation between light and dark colors. We can see a frontier line on distribution separating it into two regions. This line represents the unstable fixed point $x^*=a/(a+1)$ of the map.

 The periodic behavior of the density becomes clearer if we realize a proper partitioning of phase space. Using the fixed point of the map as a boundary we split the phase space into two regions. We call these regions $0$ and $1$, as shown in Fig. 1(b).  We observe that all points in  $0\ (1)$  are mapped to  $1\ (0)$. It is easy to verify that any point outside that region is eventually mapped to it.

Now we present an alternative procedure in order to identify the statistical periodicity for the binary pattern of the lattice.  We define the binary pattern regarding the fixed point of the map: we assign the value $1$ if  $x_n^{(i)}> x^*$ or $0$ if $x_n^{(i)}\le x^*$  \cite{PRE67} .  To avoid finite size problems, we assume periodic boundary conditions. The initial conditions are uniformly distributed on the interval $[a(1-a/2),a/2]$, in which the invariant density has non-null supports.

\section{The entropy rate}

The entropy rate is defined  \cite{PRL85} as the average number of bits to encode one additional state when we know all previous states of the system. Denoting for  $x'^{(i)}$ the binary state of the  $i-$th site  of the lattice, the entropy rate  of the system is given by
\begin{eqnarray}
\label{taxa}
h=-\sum_{i=1}^{N} p(x'^{(i)}_{n+1},x'^{(i)}_k)\ln p(x'^{(i)}_{n+1}|x'^{(i)}_k)
\end{eqnarray}
where $p(x'^{(i)}_{n+1},x'^{(i)}_k)$ is the joint probability, $ p(x'^{(i)}_{n+1}|x'^{(i)}_k)$ the conditional probability and $x'^{(i)}_k$ a shorthand notation for $(x_n^{(i)},x_{n-1}^{(i)},\cdots,x_{n-k+1}^{(i)})$. Thus, the entropy rate measures the uncertainty degree of the state at instant $n+1$ with respect to $k$ previous states.

\begin{figure}[h]
\centering
\begin{minipage}{14pc}
\includegraphics[width=15pc,angle=0,clip]{fig3.eps}
\end{minipage}\hspace{2pc}%
\begin{minipage}{14pc}
  \includegraphics[width=16pc,angle=0,clip]{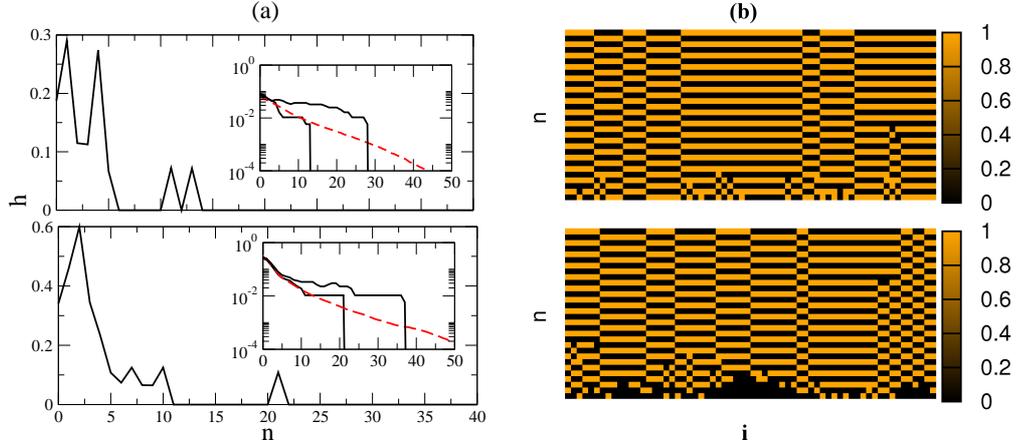}
\end{minipage} 
\caption{\label{Fig2} Entropy rate (a) and binary activity pattern (b) as a function time for $N=64$ and $\varepsilon=0.20$. The inset in (a) shows the mean entropy rate on interval $\tau=20$ for two different sets of initial conditions. The dashed curve (red online) corresponds to mean entropy rate on $1,000$ sets of initial conditions. In (b) the space (time) correspond to the horizontal (vertical) axis while the colors code the binary states. Color intense (yellow online) correspond to $1$ and dark black $0$. Figures appear at the top correspond to random initial conditions distributed on $[a(1-a/2),a/2]$  and those below on intervals $[0,1]$.}
\end{figure}

Note that, for systems whose the asymptotic pattern presents statistical periodicity with a arbitrary period $T$, we have  $p(x^{(i)}_{n+1}|x^{(i)}_{T-1})=1 \ \forall\ i$, which implies  $h=0$. Therefore, since the entropy is null only if the pattern has periodicity, this quantity can be used to calculate the locking-time. Figure \ref{Fig2} presents the entropy rate  $h(k=2)$ and the binary pattern as a function of the time for the lattice (\ref{rede}), for two different initial distributions. We can see that the entropy rate abruptly decays to zero if period-2 oscillations occurs. We also observe that there are transient states which oscillate, before to reach the asymptotic behavior. In Fig. \ref{Fig2}(a), for instance, we note a oscillation state, with period-2, between  $n=6$ and $n=10$, which coincides with $h=0$. After this interval we note a small fluctuation in the pattern, and only at $n=13$ the asymptotic state is reached. The similar behavior is observed for the case in which the initial conditions are distributed on the interval $[0,1]$ ( Fig.  \ref{Fig2} (b)).   

To ensure that the system reached the steady state we perform a temporal average for $h$ on the time interval $\tau$. In the inset of Fig. \ref{Fig2} we plot $h(\tau=20)$ for two different sets of initial conditions. We see that, in this case, $h(\tau=20)$ goes to zero only if the system reaches the asymptotic periodic state. We also note that the convergence of the pattern depends on the initial conditions. So, for a reasonable estimate of the locking-time is appropriate to consider the means on the ensemble of initial conditions. Thus, we rewrite the equation (\ref{taxa}) as follows
  \begin{eqnarray}
\langle h\rangle =-\sum^{\#} \sum^{\tau} \sum^N  p(x'^{(i)}_{n+1},x'^{(i)}_{T-1})\ln p(x'^{(i)}_{n+1}|x'^{(i)}_{T-1}),
\end{eqnarray}
where the first summation refers to initial conditions, the second one to time interval, and the last one takes into account the elements of the lattice. The dotted curve in the inset of the Fig.  \ref{Fig2} indicates $\langle h\rangle$, which is calculated with $1,000$ random initial distributions.

\section{Results}

\subsection{The locking-time}

\begin{figure}[h]
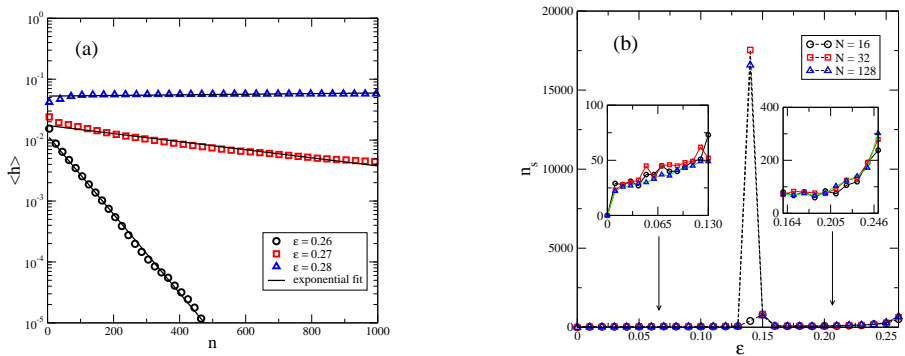
\centering
\begin{minipage}{14pc}
\includegraphics[width=12pc,angle=0,clip]{fig5.eps}
\end{minipage}\hspace{2pc}%
\begin{minipage}{14pc}
  \includegraphics[width=12pc,angle=0,clip]{fig6.eps}
\end{minipage} 
\caption{\label{Fig3} (a) Mean entropy rate as a function of the time for $N=128$ and $20,000$ sets of initial conditions. (b) Locking-time as a function of the coupling strength for $N=16,32,128$, $\tau=100$ and $n_{max}=10^7$ iterates. The initial conditions were randomly distributed on the interval $[a(1-a/2),a/2]$.}
\end{figure}

In this work, our first analysis about the influence of coupling on 
the periodic behavior was related with the locking-time, which is the 
needed time for the systems to present a permanent  behavior \cite{PRE67} .
  As the periodicity of the binary pattern implies $\langle h\rangle=0$, we initially investigated the behavior of $\langle h\rangle$ as a function 
of time for different values of $\varepsilon$ and initial distributions, in order to identify the values for the locking-time. Figure  \ref{Fig3}(a) exhibits $\langle h\rangle$ calculated on a ensemble of $20,000$ random initial distribution, with $\tau=100$, $\varepsilon=0.26$, $\varepsilon=0.27$ and $\varepsilon=0.28$. We observe that at the interval of time in the figure $\langle h\rangle$ follows an exponential law whose slope depends on $\varepsilon$, {\it i.e.}, $\langle h\rangle\propto n^{\beta(\varepsilon)}$. For $\varepsilon =0.26$ and $\varepsilon =0.27$ the coefficients are negative, $\beta(0.26)\approx -10^{-2}$ and  $\beta(0.27)\approx -10^{-3}$, while for $\varepsilon =0.28$ we obtained $\beta(0.28)\approx 10^{-4}$.

The qualitative change in the slope for the $\langle h\rangle$ plot in Fig \ref{Fig3}(a)  suggests the existence of a critical value of $\varepsilon$, above that  periodicity will no longer be observed. To go into the details in this question, we calculate the mean locking-time $n_s$ for different values of $N$. We consider a maximum time observation of $10^7$ iterations, and $10^{-6}$ as a threshold values for $\langle h\rangle$, below that  we consider  the system is in the periodic permanent state. Figure \ref{Fig3}(b) shows the dependence of $n_s$ on $\varepsilon$ for $N=16$, $32$ and $128$. Since the distribution of initial conditions respected the density interval for the map, we obtained a locking-time equals 1 with $\varepsilon=0$, for $\varepsilon>0$ the locking-time is directly proportional to $\varepsilon$. Besides, the curve presents a peak divergence in $\varepsilon\approx 0.15$ for all values of $N$, and for $\varepsilon>0.27$ we did not obtain any value for $n_s$ until $10^7$ iterations.

\begin{figure}[h]
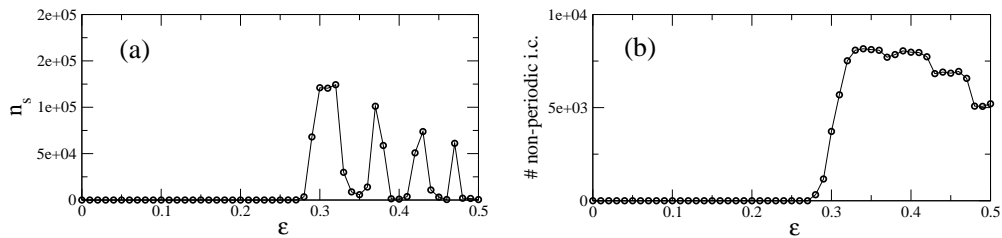

\centering 
\begin{minipage}{14pc}
\includegraphics[width=15pc,angle=0,clip]{fig8a.eps}
\end{minipage}\hspace{2pc}%
\begin{minipage}{14pc}
  \includegraphics[width=15pc,angle=0,clip]{fig8b.eps}
\end{minipage}  
\caption{\label{Fig5}  Mean locking-time (a) and number of non-periodic initial  condition sets (b) as a function of the coupling strength. The mean locking-time was weighed by distributions for which  $\langle h\rangle_\tau <10^{-6}$ in $10^7$.  We consider $N=32$, $\tau=100$ and $10.000$ random initial distributions on intervals $[a(1-a/2),a/2]$.}
\end{figure}

Although we have not found any values of $n_s$ for $\varepsilon>0.27$, we cannot assert that they do not exist. We have three possibilities: a) $n_s>10^7$ iterations; b) the period of oscillation is greater than 2; c) the binary pattern ceases to exist for $\varepsilon>0.27$. Once we determined the locking-time from an ensemble of initial distributions, some of them can take more than $10^7$ iterations to reach the stationary state, and contributed significantly for the calculation of $\langle h\rangle$. To investigate this hypothesis we verify the portion of initial distribution that satisfy $\langle h\rangle<10^{-6}$ for $n\le10^7$. Figures  \ref{Fig5}(a) and \ref{Fig5}(b) present the graphical verification of the hypothesis for a ensemble of $10,000$ initial conditions and $N=16$, the observation time  being equal to $10^7$. The mean locking-time is small -- $\mathcal{O}(10^2)$ iterations -- for $\varepsilon<0.27$, exhibits a maximum value in the order of $10^5$ iterations for $\varepsilon\in(0.27,0.34)$, and a irregular oscillation after that interval. The number of initial distribution that has not reached the periodic state until $10^7$ iterations is equals to zero when $\varepsilon<0.27$, however, for values greater than the former, nearly $70\%$ of the distributions have not converged yet. Therefore, our results suggest that the oscillation of period two ceases to be the only typical behavior if $\varepsilon>0.27$. We should pay attention to the fact that this critical value do not depends on the size of the lattice.

\begin{figure}[h]
\centering
\includegraphics[width=14pc,angle=-90,clip]{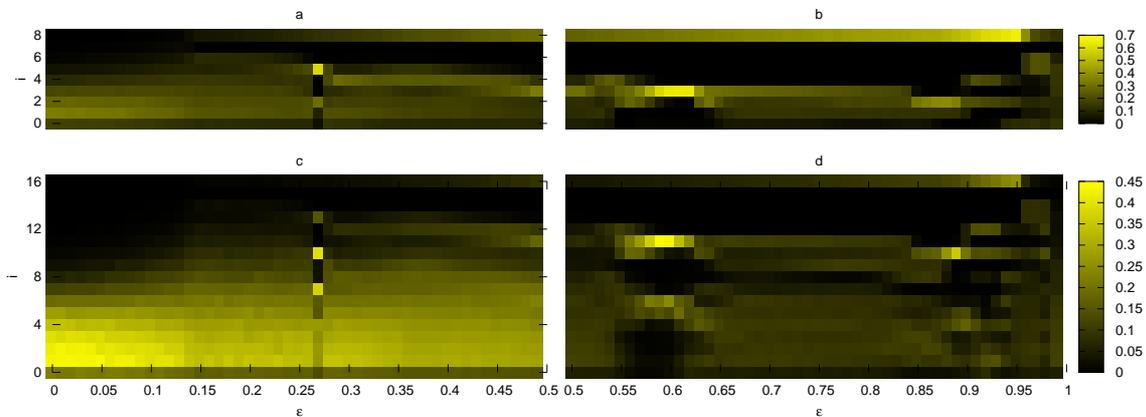}
\caption{\label{Fig4} Density of effective binary states as a function the coupling strength. (a) and (b) refers to $N=16$, (b) and (c) $N=32$. The states (coupling strength) correspond to vertical (horizontal) axis while color code the state density. Color intense (yellow online) correspond to higher density and dark color lower density. The density were weighted on $20,000$ initial distribution and $100$ iterates, considering $10^6$ of transient.}
\end{figure}

Aiming to examine the peak and the divergence of the $n_s$ in Fig. \ref{Fig3}(b), we studied the density of the effective binary states as the parameter $\varepsilon$ suffers variation. We define and label each effective binary state $s$ in relation to the number of $0$'s and $1$'s that appears in the binary pattern after a transient time, {\it i.e.}

 \begin{eqnarray}
s \equiv \left\{\begin{array}{ll}
  \frac{1}{2}|N_0-N_1| & \mbox{if $N$ is even}\\
\frac{1}{2}(|N_0-N_1|-1) & \mbox{if $N$ is odd}\end{array}\right.
\end{eqnarray}
in which, $N_0(N_1)$ is the number of $0$'s($1$'s) in the asymptotic binary pattern. From this definition follows that a lattice with $N$ elements have $\frac{N}{2}+1$ effective binary states if $N$ is even and $\frac{N+1}{2}$ states if $N$ is odd. Figure \ref{Fig4} shows $s$ for $N=16$ and $N=32$, where we used $20,000$ initial distributions e considered $100$ iterations after a transient of $10^6$. In Fig.  \ref{Fig4}(a) we see that when $\varepsilon\approx 0.15$ the density of some states $s$ abruptly varies, which occurs for the states $6$, $7$ and  $8$. More precisely, we see that while the occupation probability of the states 6, and 8 increases, for the state 7 it goes to zero. Because of this the locking-time diverges in that point. We note another abrupt change in the probability at $\varepsilon\approx0.27$, for which three new states becomes unlikely. Therefore, we attribute the divergence of time-locking to the fact that some states have almost null probability of being occupied.

\subsection{The information capacity}

In the previous section we see that the intensification of the coupling implies destruction of binary patterns. Since the multiplicity of the patterns is related to information capacity \cite{PRE67}, we have that the stronger the coupling, the lower the information capacity in the lattice. This means that the intensification of the interaction between sites decreases the uncertainty about the asymptotic states. A way to quantify such uncertainty is through the conditional entropy given by

\begin{eqnarray}
  H(x^{(i)}|y^{(i)})=-\sum_{x^{(i)}}\sum_{y^{(i)}} p(x^{(i)},y^{(i)})\ln p(x^{(i)}|y^{(i)}),
\label{ent}
\end{eqnarray} 
where $x^{(i)}$ is the binary state of the $i-$th element, $y^{(i)}$ are the binary states of the next-neighbors of $i$, $p(x^{(i)},y^{(i)})$ is the joint probability and  $p(x^{(i)}|y^{(i)})$ is the conditional probability.

The conditional probability is one that concerns the element be "1" or "0" if its neighbors, to which it is connected , are in a certain configuration. Such probability is given by

\begin{eqnarray}
 p(x^{(i)}|y^{(i)})=\frac{p(x^{(i)},y^{(i)})}{\sum_{x}p(x^{(i)},y^{(i)})} 
\label{probcond}
\end{eqnarray} 
in which the summation for $x$ is over the two possible states of $i-$th element. 

\begin{figure}[h]
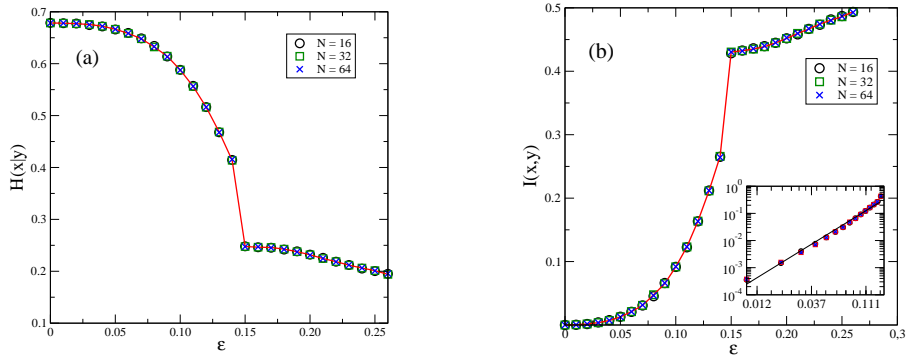

\centering 
\begin{minipage}{14pc}
\includegraphics[width=12pc,angle=0,clip]{fig13.eps}
\end{minipage}\hspace{2pc}%
\begin{minipage}{14pc}
  \includegraphics[width=12pc,angle=0,clip]{fig14-new.eps}
\end{minipage} 
\caption{\label{Fig8} Conditional entropy or entropy per element (a) and mutual information (b) as a function of coupling strength. These quantities were evaluated  on the permanent binary pattern of $20,000$ initial condition sets. We consider the CML on permanent binary pattern when $h<10^{-12}$.}
\end{figure}

Using Eqs.  (\ref{probcond}) and (\ref{ent}) we evaluate the conditional entropy of the binary pattern which corresponds to the lattice  (\ref{rede}) as a function of $\varepsilon$ on the interval $[0.00,0.27]$. Figure \ref{Fig8}(a) shows the dependence of  $H(x^{(i)}|y^{(i)})$ on $\varepsilon$ for different values of $N$. When $\varepsilon=0.0$ the entropy is equal to $\ln 2$. Until $\varepsilon\approx 0.15$, $H(x^{(i)}|y^{(i)})$ presents a more pronounced decay than those for $\varepsilon>0.15$. The curves  For the three different values of $N$ the curves coincide. This indicates that the conditional entropy do not depends on the size of the lattice.

Conditional entropy, Eq. (\ref{ent}), also can be used to obtain the mutual information. This quantity corresponds to information obtained about the state of one element by knowing the states of its neighbors. Thus, we have
\begin{eqnarray}
I(x,y_i)&=&H(x)-H(x|y_i)
\label{Eq_IM}
\end{eqnarray} 
where $H(x)\approx \ln(2)$ is the entropy of $x$. 

In Fig. \ref{Fig8}(b) we show the behavior of $I(x,y_i)$  as a function of $\varepsilon$. We note that the mutual information increases with the coupling parameter. The fit curve for $\varepsilon\in(0,0.15)$ (inset) shows that it follow a power-law,{\it i.e.} $I\propto\varepsilon^\alpha$, being $\alpha\approx 2.6$.  As for the entropy, an abrupt changes in behavior of $I(x,y_i)$ occurs at $\varepsilon\approx 0.15$, which the first maximum of the locking-time and where one effective binary state becomes unlikely.
With this variation the mutual information (conditional entropy) assumes a larger (smaller) value from its mean value and therefore the state of an arbitrarily chosen element is known for his neighborhood more than $50\%$ the uncoupled case one. 

On the multiplicity of binary patterns in response to the variability of initial distributions, our results show that it becomes smaller increasing the interaction between network sites. For $\varepsilon=0$, for example, about $2^{N H/ \ln 2}\approx 2^N$ while for $\varepsilon=0.25$ it is reduced to approximately $2^{N/3}$ possible patterns. Therefore, both characteristics of neural networks, optimal capacity information and smaller locking-time, are verified when the network elements interact only weakly.

\section{Conclusion}
We studied the information capacity of a coupled map lattice with statistical periodicity considering the multiplicity of the patterns into periodic state. We propose a method to determine the locking-time of systems with periodic states and we apply it to CML under study.  We found that periodic behavior and oscillation period of the uncoupled map are preserved by CML for weak coupling strength. However, we observed that the average locking-time and the number of initial patterns that not achieve the periodic state is large for strong coupling. Our results, supported by the entropy rate and conditional entropy, suggest that the information capacity decreases and the locking-time increase  as the interaction between the sites of the CML becomes more intense. Therefore, if the interest is to study neural networks using coupled map lattices, weak coupling are required. On the other hand, if the objective is more efficient communication, it should be used strong coupling strength.

\ack
This work was made possible through partial financial support from the following Brazilian research agencies: Funda\c c\~ao Arauc\'aria, CAPES and CNPQ.

\section*{References}

\bibliography{iopart-num}

\end{document}